\newcommand{\CLASSINPUTinnersidemargin}{18mm}
\newcommand{\CLASSINPUToutersidemargin}{12mm}
\newcommand{\CLASSINPUTtoptextmargin}{20mm}
\newcommand{\CLASSINPUTbottomtextmargin}{25mm}
\documentclass[10pt,conference,twoside,a4paper]{IEEEtran}

\usepackage[utf8]{inputenc}

\usepackage{times}

\usepackage{abstract}
\renewcommand{\abstractname}{}    

\usepackage{graphicx}
\usepackage{subfigure}
\usepackage{svg}
\DeclareGraphicsExtensions{.png,.eps,.ps,.pdf,.svg}

\usepackage{fancyhdr}
\usepackage{kantlipsum}

\usepackage{url}

\usepackage{tabularx}

\usepackage[left=2cm,top=2.5cm,right=2cm,bottom=2.5cm]{geometry}

\usepackage[]{color}

\usepackage{eso-pic}

\begin{document}
\AddToShipoutPictureBG*{%
  \AtPageUpperLeft{%
    \setlength\unitlength{1in}%
 	\hspace{2cm}
 	 	\makebox(0,-2)[l]{
			\begin{tabular}{l r} 
			\multicolumn{1}{p{12cm}}{\vspace{-0.3cm}\includegraphics[scale=0.13]{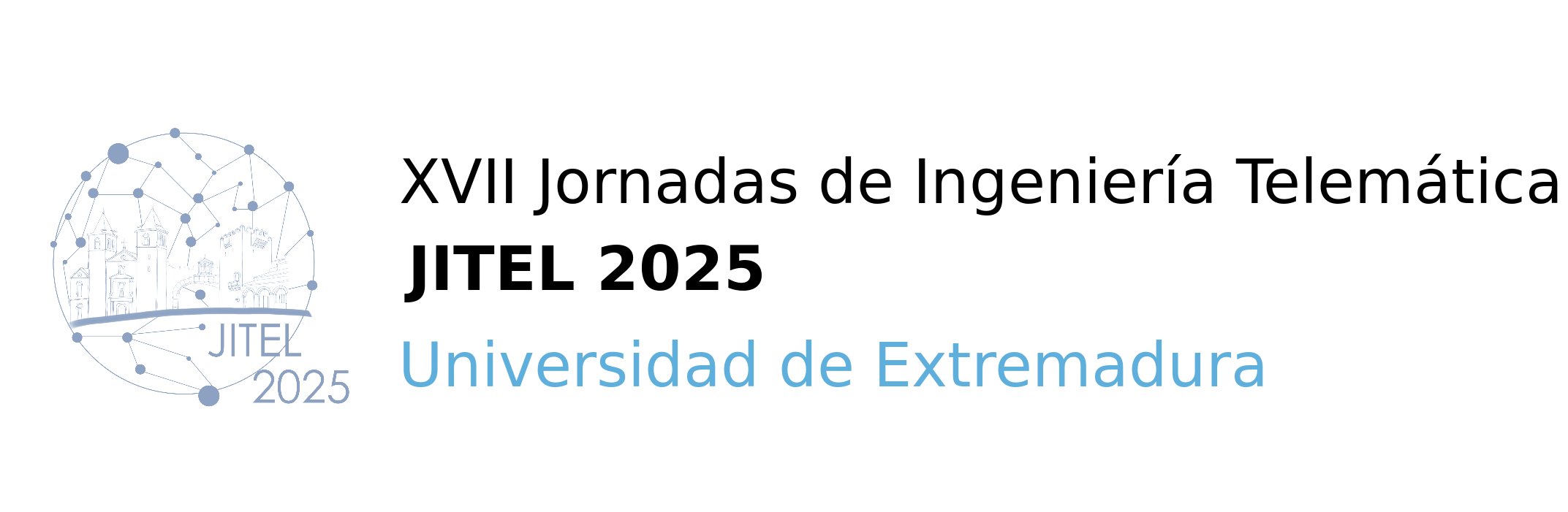}} & \multicolumn{1}{p{4cm}}{\raggedleft\small\usefont{T1}{phv}{m}{it} Actas de las XVII Jornadas de Ingeniería Telemática\\ (JITEL 2025),\\ Cáceres (España), \\12-14 de noviembre de 2025. \vspace{0.2cm} \\ISBN} \tabularnewline 
			\end{tabular}
 }
}}

\AddToShipoutPictureBG*{%
  \AtPageLowerLeft{%
    \setlength\unitlength{1in}%
    \hspace*{\dimexpr0.5\paperwidth\relax}
    \makebox(0,1.3)[c]{\footnotesize\usefont{T1}{phv}{m}{} This work is licensed under a \underline{\textcolor{blue}{Creative Commons 4.0 International License}} (CC BY-NC-ND 4.0)}%

}}

\title{\vspace{3cm}Propuesta de implementación de catálogos federados para espacios de datos sobre DataHub }

\author{\IEEEauthorblockA{Carlos Aparicio de Santiago, Pablo Viñuales Esquinas, Irene Plaza Ortiz,\\ Andres Munoz-Arcentales, Gabriel Huecas, Joaquín Salvachúa, Enrique Barra}
\IEEEauthorblockA{Information Processing and Telecommunications Center, 
Universidad Politécnica de Madrid.\\
ETSI Telecomunicación. Avda. Complutense 30, Madrid 28040, Spain.\\
ce.aparicio@upm.es, p.vinualese@upm.es, irene.plaza.ortiz@alumnos.upm.es, joseandres.munoz@upm.es,\\
gabriel.huecas@upm.es, joaquin.salvachua@upm.es, enrique.barra@upm.es}
}

\maketitle

\begin{abstract}
\textbf{
En la era digital, los espacios de datos emergen como ecosistemas clave para el intercambio seguro y controlado de información entre participantes. Para ello, son esenciales componentes como los catálogos de metadatos y los conectores de espacios de datos. Este documento propone una implementación y solución de integración para ambas piezas, considerando las directrices de estandarización en formatos de datos, metadatos y protocolos, lo que garantiza la interoperabilidad. \newline \newline
Se presenta una solución híbrida: DataHub se utiliza como catálogo federado para una gestión robusta de metadatos, aprovechando sus capacidades avanzadas de ingesta, gobernanza y linaje. Por otro lado, la implementación propia, Rainbow Catalog, gestiona las políticas ODRL de acceso y uso. Esta integración permite consultar \textit{datasets} de DataHub y asociarles políticas ODRL, facilitando los flujos de negociación y transferencia definidos por el Dataspace Protocol. El resultado es un sistema que combina la potencia de DataHub para la catalogación a gran escala con la gestión de políticas del conector, clave para la soberanía y confianza en los espacios de datos.
\newline
}
\end{abstract}


\section{\uppercase{Introducción}}
La digitalización acelerada de las últimas décadas ha transformado profundamente todos los sectores de la sociedad, incluida la economía. Esto ha impulsado la consolidación de los datos como recurso esencial, tanto en contextos profesionales como en la vida cotidiana.

En respuesta a esta transformación, a principios de los años 2000 surgió el concepto de Open Data, promovido por organismos como la Unión Europea, con el objetivo de facilitar el acceso público a la información y reforzar la transparencia institucional \cite{attard2015systematic}. Desde entonces, se han creado numerosos portales de datos abiertos \cite{schmidt2022practices}, que actúan como repositorios con funciones de búsqueda y filtrado \cite{wibowo2023systematic}. Sin  embargo, la adopción de estos portales ha estado frenada por obstáculos como la publicación de datos incompletos, formatos no interoperables, metadatos deficientes y la escasa participación del sector privado \cite{umbrich2015quality, beno2017open}.

Los espacios de datos surgen como una alternativa a estos entornos. Dichos espacios de datos pretenden la generación de valor alrededor del dato mediante su compartición en un entorno de soberanía, confianza digital y seguridad \cite{scerri2022common, bacco2024data}. Estos espacios de datos están construidos sobre principios de descentralización, interoperabilidad, gobernanza y transparencia \cite{martella2025designing}. 

Para cumplir estos requisitos, se propone el uso de tecnologías emergentes como la identidad autosoberana (SSI) \cite{odrl22}\cite{menendez2025next}, apoyada en estándares del W3C como los \textit{Decentralized Identifiers} (DID) y las \textit{Verifiable Credentials} (VC). Asimismo, se emplean protocolos comunes como el \textit{Dataspace Protocol} \cite{idsa2024}. Para definir políticas de acceso y uso, se recurre al modelo ODRL 2.2 \cite{dam2023policy}, y para la descripción de recursos y gestión de catálogos, al vocabulario DCAT3 \cite{dcat3}\cite{hauff2024fairness}; ambos son recomendaciones oficiales del W3C. Además, los espacios de datos requieren infraestructuras avanzadas que habiliten no solo el intercambio de datos, sino también servicios de análisis, inferencia y compartición segura de datos personales, así como aplicación de políticas de control de acceso y uso que estén vinculadas a los activos digitales.

A diferencia de los entornos de datos abiertos, los espacios de datos no implican acceso libre e ilimitado a la información. En este nuevo paradigma, el acceso está sujeto a acuerdos explícitos entre proveedores y consumidores de datos, lo que requiere mecanismos de negociación de condiciones de uso. Una vez alcanzado un acuerdo, se formaliza mediante un contrato que define los términos de la transferencia. Este proceso es gestionado por un componente técnico denominado conector de espacios de datos (\textit{Dataspace Connector}) \cite{dam2024surveydataspaceconnectorimplementations}.  

Adicionalmente, el ecosistema requiere de un catálogo de metadatos, encargado de almacenar, publicar y facilitar el acceso a los conjuntos de datos disponibles. También requiere, a su vez, la simplificación y automatización de ingesta de metadatos y conexión con sistemas finales de datos. Este catálogo actúa como punto de descubrimiento, permitiendo a los proveedores registrar sus recursos y a los consumidores localizarlos, explorarlos o acceder a ellos, mediante funcionalidades avanzadas de búsqueda y agregación. Además, permite definir las políticas de uso y acceso asociadas a dichos servicios de datos para tener un entorno de gobernanza. \cite{morejon2025exploring} 

Este artículo propone una arquitectura técnica para un catálogo federado en espacios de datos, integrando la plataforma DataHub con el conector interoperable Rainbow. El modelo híbrido resultante aprovecha DataHub para la gestión avanzada de metadatos y Rainbow para el control de acceso mediante políticas ODRL. Se analiza cómo el uso de estándares como DCAT3 garantiza la interoperabilidad del ecosistema, abordando así los principales retos de la federación y gobernanza de datos.

\section{\uppercase{Catálogo fuente y Catálogo federado}}
Los catálogos constituyen un componente fundamental en la arquitectura de los espacios de datos, al encargarse de la gestión y exposición de conjuntos de datos enriquecidos con metadatos. Actúan como punto de encuentro entre proveedores, que publican sus recursos, y consumidores, que acceden a ellos, los consultan y los tratan bajo determinadas condiciones.

Con todo, es posible distinguir dentro de los espacios de datos dos tipos complementarios de catálogos: catálogos fuente y catálogos federados \cite{theissen2024towards}.

En primer lugar, los catálogos fuente, también llamados catálogos de origen, son gestionados por los propios proveedores de datos. En ellos se publican \textit{datasets}, acompañados de metadatos que describen características clave (por ejemplo, el formato, la frecuencia de actualización o las condiciones de acceso). Esta información es esencial, ya que consigue enriquecer los conjuntos de datos de modo que los consumidores puedan evaluar la relevancia y adecuación de los recursos antes de acceder a los mismos.

A medida que crece el número de proveedores y catálogos individuales, se hace necesaria la existencia de catálogos federados. Estas plataformas permiten integrar y agrupar los recursos publicados en múltiples catálogos fuente, ofreciendo a los usuarios un punto de entrada unificado al ecosistema. Esto no solo facilita la búsqueda y el descubrimiento de datos, sino que también amplía el alcance de los proveedores y fomenta un entorno más accesible y eficiente. \cite{jahnke2024federated}

\section{\uppercase{Conector de espacios de datos}}
Aunque todavía no existe una definición plenamente consensuada a nivel de arquitectura de lo que constituye un Espacio de Datos, hay esfuerzos por parte de organismos de estandarización y grupos de trabajo de UNE, CEN-CENELEC para definirla y se están haciendo contribuciones a los estándares W3C, IETF y ETSI en la misma línea. No obstante, la mayoría de enfoques coinciden en señalar al conector como uno de sus componentes esenciales. Esta pieza cumple un papel central en la arquitectura del ecosistema, ya que actúa como el mecanismo que habilita la interoperabilidad entre participantes. En concreto, el conector es responsable de gestionar los procesos de negociación de políticas de acceso y uso, así como de transferencia de datos, garantizando que estas operaciones se realicen bajo marcos comunes y estandarizados. \cite{dam2024surveydataspaceconnectorimplementations}

Además de garantizar la interoperabilidad entre actores, los conectores deben incorporar mecanismos de autenticación y autorización que se alineen con los principios de identidad soberana (SSI). Para ello, se pueden emplear sistemas basados en \textit{Verifiable Credentials} y \textit{Verifiable Presentations}, implementados a través de protocolos como OpenID4VCI \cite{openid4vci}, OpenID4VP \cite{openid4vp}, DCP \cite{dcp} o GNAP4VP \cite{menendez2025next}. Estos protocolos permiten verificar la identidad de los participantes controlando las credenciales presentadas, garantizando un marco de confianza digital.  

Los conectores de espacios de datos manejan los cuatro tipos de actores ilustrados en la Figura 1:  
\begin{figure*}[htbp] 
\centerline{
\includegraphics[width=\textwidth]{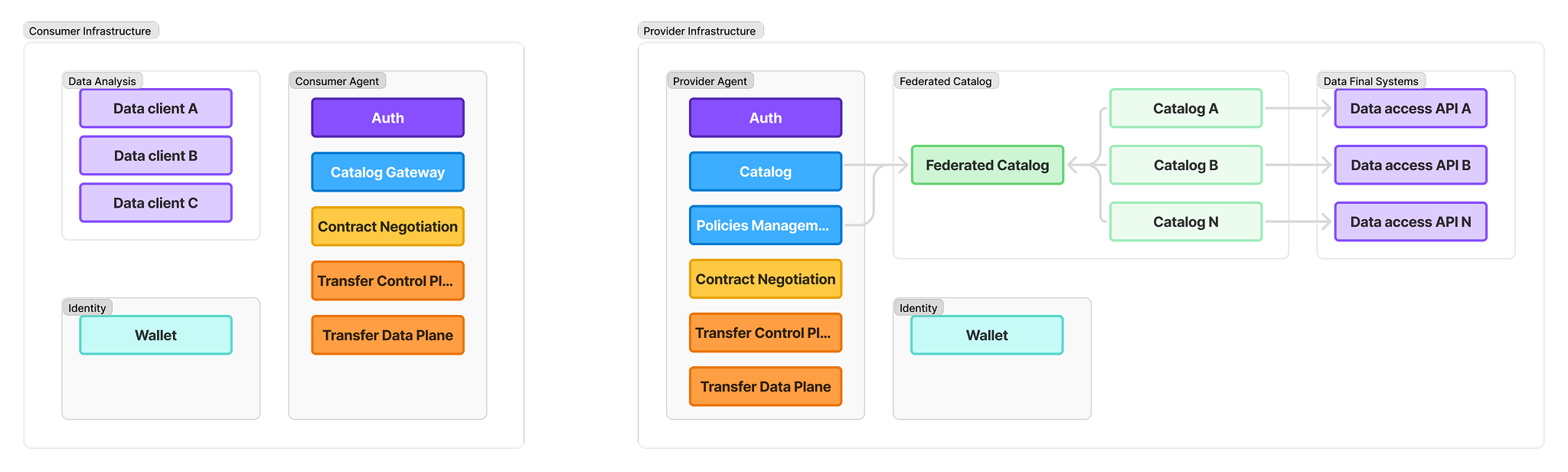}
}
\caption{Diagrama de contexto de un conector de espacios de datos.}
\label{fig:protocols}
\end{figure*}
\begin{itemize}
    \item \textit{Dataspace Provider}: componente software responsable de exponer el catálogo de metadatos, gestionar el acceso a los datos, y ejecutar los procesos de negociación y transferencia según estándares de interoperabilidad. Además, se conecta con los sistemas finales que alojan los datos reales.
    \item \textit{Dataspace Consumer}: contraparte del proveedor, implementa los protocolos necesarios para negociar y consumir los datos conforme a las políticas establecidas.
    \item \textit{Sistemas finales de datos}: fuentes donde residen los datos, accesibles únicamente por el \textit{Dataspace Provider}.
    \item \textit{Clientes de datos}: procesos o aplicaciones que acceden a los datos a través del Dataspace Consumer.
\end{itemize}
Además de los actores funcionales, el ecosistema también contempla entidades de gobernanza distribuida y autoridades de confianza digital, que no se abordan en este trabajo.
En términos de arquitectura, el cliente de datos y el \textit{Dataspace Consumer} operan dentro de una misma infraestructura y bajo un marco de gobernanza común, al igual que el \textit{Dataspace Provider} y el sistema final de datos. Estos dos dominios —aislados entre sí— se comunican mediante el conector, que orquesta los flujos de autenticación, negociación de políticas y transferencia de datos. En conjunto, el \textit{Dataspace Provider}, el \textit{Dataspace Consumer} y sus mecanismos de interacción constituyen lo que se conoce como un conector de espacio de datos.

\section{\uppercase{Papel del catálogo en el conector de Espacio de Datos}}
Como se explicó en el epígrafe anterior, uno de los requisitos del actor \textit{Dataspace Provider} es exponer un catálogo de metadatos. Esto resulta esencial, ya que habilita los flujos clave del ecosistema: la negociación de contratos y la negociación de transferencia de datos. Ambos procesos están definidos en el \textit{Dataspace Protocol} y requieren que los recursos y sus políticas asociadas estén previamente publicados en el catálogo.

\subsection{Flujo de negociación de contrato} 
El proceso de negociación de contrato se modela como una máquina de estados basada en el intercambio de mensajes entre el \textit{Dataspace Consumer} y el \textit{Dataspace Provider} (como se muestra en la Figura 2). Este flujo puede iniciarse por cualquiera de las partes, mediante mensajes como \textit{ContractRequestMessage} o \textit{ContractOfferMessage}. A través de sucesivos intercambios, se alcanza un acuerdo formalizado o contrato (también denominado \textit{Agreement}), que definirá las condiciones de acceso y uso del recurso.

\begin{figure}[htbp] 
\centerline{
\includegraphics[width=\columnwidth]{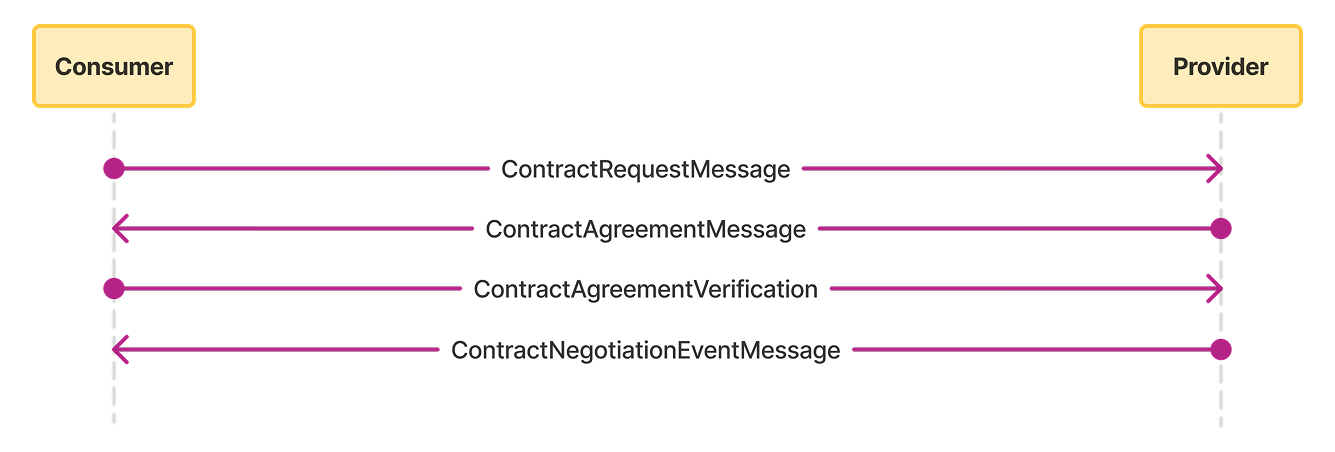}
}
\caption{Flujo de negociación de contrato en relación a ODRL}
\label{fig:protocols}
\end{figure}

Estos mensajes intercambiados durante la negociación del contrato contienen referencias a políticas expresadas en ODRL. ODRL (\textit{Open Digital Rights Language}) es un lenguaje de expresión de políticas que permite describir condiciones de acceso y uso de datos con un alto nivel de granularidad. Sin profundizar en el modelo, es importante señalar que cada política hace referencia explícita al recurso al que se aplica, el cual debe estar previamente registrado en el catálogo de metadatos.

De este modo, la existencia del catálogo no solo es necesaria para la identificación del recurso, sino también para validar la coherencia entre lo negociado y lo publicado. El contrato resultante, por tanto, queda vinculado al recurso definido en el catálogo, así como a los participantes del acuerdo: el proveedor (\textit{Dataspace Provider}) y el consumidor (\textit{Dataspace Consumer}). Actualmente, existen esfuerzos dedicados a mejorar y extender el lenguaje ODRL 2.2 para que se adapte a las necesidades específicas de los espacios de datos \cite{plaza2025authentication} y se está contribuyendo en la especificación de ODRL 3.0.

\subsection{Flujo de negociación de transferencia}
El \textit{Transfer Process Protocol}, representado en la Figura 3, describe la negociación para la transferencia efectiva del recurso. Este flujo también se estructura como una máquina de estados. En este caso, el \textit{Dataspace Consumer} inicia el proceso con un \textit{TransferRequestMessage}, que hace referencia al identificador del contrato previamente alcanzado, y al tipo de distribución deseada.
\begin{figure}[htbp] 
\centerline{
\includegraphics[width=\columnwidth]{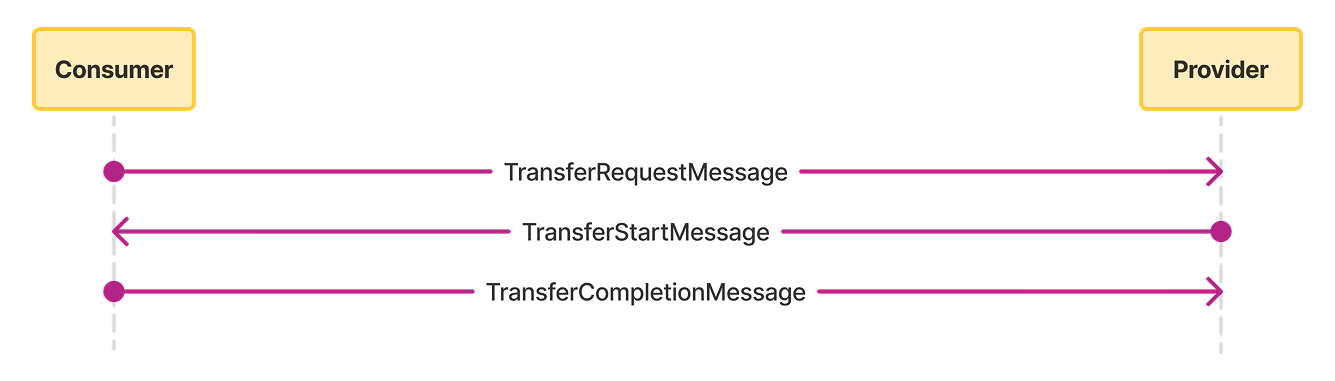}
}
\caption{Flujo de negociación de transferencia en relación a ODRL.}
\label{fig:protocols}
\end{figure}
Una vez verificado, el sistema genera un punto de acceso a los datos, utilizando la información registrada en el catálogo. 

\section{\uppercase{Captura de requisitos}}
La definición de requisitos para un catálogo federado en el contexto de los espacios de datos ha sido abordada ampliamente en la literatura reciente \cite{jahnke2023data}. Existen, además, referencias relevantes en materia de estandarización de metadatos, gestión del linaje de datos y otros aspectos técnicos derivados de experiencias prácticas de implementación e integración. Dichos requisitos se definen en la Tabla I.

\begin{table}[htbp] 
\centering
\caption{Tabla de Requisitos.} 
\label{tab:requisitos} 
\renewcommand{\arraystretch}{1.25} 
\begin{tabularx}{\columnwidth}{c | X} 
\hline
\hline
R1 & Capacidad de generar metadatos desde sistemas federados con esquemas unificados. \cite{li2016interoperability} \\
R2 & Capacidad de resolver información sobre los metadatos en los flujos de negociación de transferencia y de políticas. \cite{li2016interoperability} \\
R3 & Capacidad del sistema de metadatos de poder almacenar y gestionar políticas de acceso y uso, así como la capacidad de indexarlas. \\
R4 & Definición de metadatos de manera interoperable con estándares internacionales reconocidos. \cite{brechtel2023challenges}\\
R5 & Definición de políticas con estándares internacionales reconocidos. \cite{noardo2024standards} \\
R6 & Sistemas automatizados para ingestión de metadatos desde las fuentes de datos o desde sistemas federados. Facilidad de integración con sistemas propios de ingeniería de datos. \\
R7 & Capacidad de generar información sobre el linaje de los datos y la calidad del dato. \cite{altendeitering2022data} \\
R8 & Capacidad de gestión de búsqueda y descubrimiento de \textit{datasets}. \\
\hline
\hline
\end{tabularx}
\end{table}

\section{\uppercase{Implementación propuesta - Rainbow Catalog }}
Como primera aproximación técnica a los requisitos identificados, se ha generado un servicio propio de catálogo de metadatos integrado en la implementación del conector de espacios de datos. Este desarrollo, denominado Rainbow, se ha centrado especialmente en garantizar la interoperabilidad, adoptando como base el vocabulario estandarizado DCAT3 para la descripción de metadatos, y permitiendo la asociación de políticas de acceso mediante el modelo ODRL 2.2, conforme a sus especificaciones. Cabe señalar que este enfoque es coherente con las directrices del propio \textit{Dataspace Protocol}, que adopta DCAT3 en su definición del \textit{Catalog Protocol}. El desarrollo de Rainbow es público y el repositorio se encuentra en el siguiente enlace: \url{https://github.com/EunomiaUPM/rainbow}

Una de las principales ventajas de esta solución es el cumplimiento de los requisitos de interoperabilidad, gracias a la adopción de estándares. Asimismo, al tratarse de una implementación desarrollada como parte integral del conector, se consigue una integración nativa con los flujos de negociación de contrato y de transferencia de datos.

Otro aspecto destacable es la capacidad de definir y vincular políticas ODRL directamente a los metadatos del catálogo, lo que permite una gestión coherente del control de acceso y uso desde la propia estructura del sistema.

No obstante, esta implementación presenta algunas limitaciones. En particular, la funcionalidad de descubrimiento de datos no está suficientemente cubierta, ya que actualmente no incorpora mecanismos avanzados de búsqueda, filtrado o enriquecimiento semántico de metadatos. Tampoco contempla la automatización del proceso de ingestión mediante sistemas de extracción de metadatos, lo que implica una alta carga operativa para los administradores encargados de mantener actualizado el catálogo.

Por último, en contextos donde se requiere gobernanza distribuida o se manejan grandes volúmenes de datos —como en organizaciones complejas o administraciones públicas—, es imprescindible contar con catálogos federados. En su estado actual, Rainbow no da soporte a este tipo de despliegue, lo que limita su aplicabilidad en escenarios más exigentes.

\section{\uppercase{Uso de DataHub para integración de fuentes de datos }}
Dada la importancia de sistemas federados y la limitación de Rainbow exploramos Datahub. DataHub es una plataforma moderna de metadatos. 

Una de las principales ventajas de la plataforma es la alta capacidad que tiene para integrarse con otras fuentes, desde las cuales se ingestan los datos. Estas fuentes pueden ser plataformas para la gestión de bases de datos relacionales o fuentes especializadas en procesos de extracción, transformación y carga. También admite la integración con sistemas de procesamiento de datos en tiempo real. Otra ventaja relevante que es que actualmente tiene una amplia adopción en sistemas en producción. 

Internamente, DataHub organiza los metadatos como un grafo de entidades interconectadas. Esta estructura no solo potencia su motor de búsqueda y descubrimiento de recursos, sino que también permite visualizar el linaje de los datos de forma robusta, una capacidad clave para la gobernanza. También permite la inclusión de metadatos, sistemas de búsquedas basadas en términos clave, gestión de dominios, ect. 

DataHub, no obstante, presenta ciertas limitaciones para su integración en espacios de datos, especialmente en términos de interoperabilidad y gestión de políticas. 

Una primera limitación se enfoca en que las propiedades de los \textit{datasets} se estructuran de forma no estandarizada, siendo meros atributos organizados como pares de clave-valor. Esto implica que para que se adhiera a estándares como DCAT sea preciso un control en en los sistemas ETL, lo que implica un paso innecesario de complejidad. 

La segunda, y más importante para este trabajo, es que DataHub carece de soporte nativo para la gestión de políticas de acceso y uso. La plataforma no permite definir ni asociar políticas expresadas en lenguajes estándar como ODRL, un requisito esencial para la soberanía y la confianza en los espacios de datos. Estas carencias motivan la necesidad de una solución híbrida.

\section{\uppercase{Comparativa entre DataHub Y Rainbow Catalog }}

\begin{table*}[h!] 
\centering
\caption{Tabla de comparativa de soluciones.} 
\label{tab:requisitos} 
\renewcommand{\arraystretch}{1.25} 
\begin{tabularx}{\textwidth}{l | X | X} 
\hline
\hline
\textbf{Característica} & \textbf{Rainbow Catalog} & \textbf{DataHub} \\ 
\hline
Función principal & Sistema de catálogo de metadatos. & Plataforma avanzada de gestión y descubrimiento de metadatos, enfocada en gobernanza y catálogo de datos moderno. \\
\hline
Modelo de Metadatos & Capacidad de resolver información sobre los metadatos en los flujos de negociación de transferencia y de políticas. & Modelo interno propio: DataHub Entities Model. \\
\hline
API & Capacidad del sistema de metadatos de poder almacenar y gestionar políticas de acceso y uso, así como la capacidad de indexarlas. & API estable (usualmente GraphQL), bien documentada y ampliamente utilizada. \\
\hline
Interfaz Gráfica (GUI) & Actualmente en desarrollo. & GUI muy completa y probada, uno de sus puntos fuertes para la exploración y gestión de metadatos.\\
\hline
Integraciones & Definición de políticas con estándares internacionales reconocidos. & Integración nativa con una vasta gama de sistemas (Bases de datos, Data Lakes, Warehouses, BI, ETL) para extracción automática de metadatos (ingesta).\\
\hline
Políticas & Soporte nativo para políticas ODRL 2.2 (a nivel de \textit{Catalog}, \textit{dataset}, Distribution, DataService). Roadmap para gestión de herencia. & Foco en gobernanza mediante etiquetado (tagging) flexible, dominios, glosarios y propietarios. No soporta ODRL nativamente. \\
\hline
Versionado/Linaje & Roadmap para versionado (DCAT3) y linaje (Relations). & Capacidades robustas de linaje de datos (visualizable en la GUI) y versionado implícito de metadatos. \\
\hline
Notificaciones & Sistema Pub/Sub para notificar cambios de estado (altas, modificaciones, bajas). & Emite eventos de cambio de metadatos, usualmente vía Kafka, para integración con otros sistemas. \\
\hline
Estándares/Compliance & Enfocado en DCAT3. Roadmap para DPROD, VCs (Verifiable Credentials) y Gaia-X. & Ampliamente adoptado en la industria, pero no específicamente diseñado alrededor de estándares como DCAT3 o Gaia-X de forma nativa. \\
\hline
Desarrollo/Comunidad & Parte del proyecto Rainbow. Roadmap activo & Proyecto Open Source muy activo y popular, con una gran comunidad y desarrollo constante. Ampliamente utilizado en la industria \\

\hline
\hline
\end{tabularx}
\end{table*}
Como se puede ver en la Tabla II, Rainbow Catalog está fuertemente dirigido a adherirse a estándares específicos como DCAT3 y ODRL 2.2, enfocándose en entornos que requieran interoperabilidad basada en estos estándares como es el caso de los espacios de datos. Por otro lado, DataHub es una solución más generalista y madura, destacando por su interfaz gráfica avanzada, sus amplias capacidades de integración para la ingesta automática de metadatos desde diversas fuentes, y su flexibilidad para la gobernanza de datos mediante un modelo propio y funcionalidades como el etiquetado y el linaje. Es una opción muy popular en la industria para construir un catálogo de datos federado y moderno.

A la luz de esta comparativa, concluimos que una buena opción puede ser una integración por capas, donde se haga una gestión de metadatos en DataHub aprovechando la capacidad de gestión de los mismos y las integraciones que tiene, y por otro lado trabajar en la gestión de políticas ODRL en Rainbow Catalog.

\section{\uppercase{Integración del catálogo y el conector de espacios de datos }}

En este apartado se describe la arquitectura y el flujo de integración del catálogo federado de DataHub con el conector de espacios de datos Rainbow.

Para aportar un caso de uso significativo y validar la implementación, se ha construido un escenario práctico que federa dos catálogos a partir de fuentes de datos abiertas utilizando DataHub, Apache Airflow y Apache NiFi.  Adicionalmente, se ha habilitado la gestión de políticas ODRL sobre los \textit{datasets} de ese catálogo desde Rainbow, de modo que se alcance gobernanza descentralizada, requisito esencial en un espacio de datos.

\subsection{Flujo de integración}

El flujo extremo a extremo que implementamos se organiza en cinco etapas:

1. Ingesta de metadatos desde catálogos individuales: La extracción y normalización de metadatos de cada catálogo de origen se realiza con Apache NiFi. NiFi permite definir grafos de procesamiento que cubren tanto la extracción desde las fuentes como las transformaciones necesarias. En este paso es crítico conformar los metadatos a DCAT. Una vez procesados, se cargan en DataHub.
    
2. Ingesta y federación: Para consolidar varios catálogos, aprovechamos las capacidades nativas de ingesta de DataHub, incluida la integración entre diferentes catálogos. La orquestación de procesos, frecuencias y monitorización se resuelve con Apache Airflow. El resultado es un catálogo federado que agrega y expone de forma unificada los metadatos provenientes de múltiples dominios.
    
3. Exposición de \textit{datasets} y catálogos a los participantes del espacio de datos: DataHub ofrece un \textit{endpoint} GraphQL para aplicaciones de terceros sobre su modelo interno de entidades. Sin embargo, en entornos de espacios de datos se exige una interfaz neutra e interoperable para consulta de metadatos. El \textit{Dataspace Protocol}, a través de su \textit{Catalog Protocol} y el uso de DCAT, define precisamente ese punto de acceso estandarizado expuesto por el \textit{Dataspace Provider}.
    
Por tanto, incorporamos una instancia puente que actúa de \textit{proxy} entre el catálogo interno de DataHub y el \textit{endpoint} DCAT requerido en el proveedor del espacio de datos. En la práctica, Rainbow dispone de un servicio de gestión de metadatos y una fachada que consulta DataHub y publica el punto de entrada DCAT conforme al \textit{Dataspace Protocol}, de modo que consumidores y entidades federadoras externas no necesitan conocer detalles internos de DataHub.
    
4. Gestión de políticas ODRL aplicadas a \textit{datasets} del catálogo federado: Dado que Rainbow está diseñado de manera nativa para manejar políticas ODRL, lo usamos como sistema de gestión. El ciclo de políticas incluiría: 

\begin{itemize}
    \item Descubrimiento: desde Rainbow se listan y seleccionan los recursos del catálogo federado (vía fachada a DataHub).
    \item Asociación: se crean políticas ODRL y se asocian al \textit{target} que identifica de forma única al \textit{dataset} federado.
    \item Persistencia: las políticas se guardan en Rainbow, incluyendo metadatos de auditoría y versionado lógico cuando aplica.
    \item Resolución: Rainbow expone resolución, búsqueda y aplicación de políticas.
\end{itemize} 
    
5. Uso de políticas en negociación de contrato y transferencia: Con el catálogo DCAT y las políticas ODRL disponibles, los participantes ya pueden iniciar las negociaciones de contrato, y ya se puede definir un sistema de control de políticas de acceso y uso asociado.

\subsection{Arquitectura de integración} 

La arquitectura adopta una topología en capas con acoplamiento débil, definidas en Figuras 4 y 5:

\begin{itemize}
\item Capa de metadatos (DataHub)
Persistenacia de metadatos: ingesta, catalogación, linaje y descubrimiento.
\item Capa de políticas y contratos (Rainbow)
Implementa la gestión de políticas ODRL, la exposición DCAT y los flujos de negociación del \textit{Dataspace Protocol}. Esta capa es la que habilita la soberanía y el control de uso.
\item Capa de integración (Fachada en Rainbow)
Componente clave que desacopla DataHub y Rainbow. Dicha capa actúa como: a) Cliente de la API de DataHub, y b) Mapeo semántico entre el modelo interno de DataHub y representaciones DCAT para publicación en el \textit{endpoint} del \textit{Dataspace Provider}. 
\end{itemize}

Esta separación permite evolucionar cada pieza de forma independiente: DataHub puede actualizar la información y metadatos; Rainbow puede evolucionar su modelo de políticas, herencias y mecanismos de evaluación.

De aquí se derivan dos flujos de trabajo: Un primer flujo de consulta de metadatos, y un segundo de negociación, persistencia y ejecución de políticas.

\begin{figure*}[htbp] 
\centerline{
\includegraphics[width=0.66\textwidth]{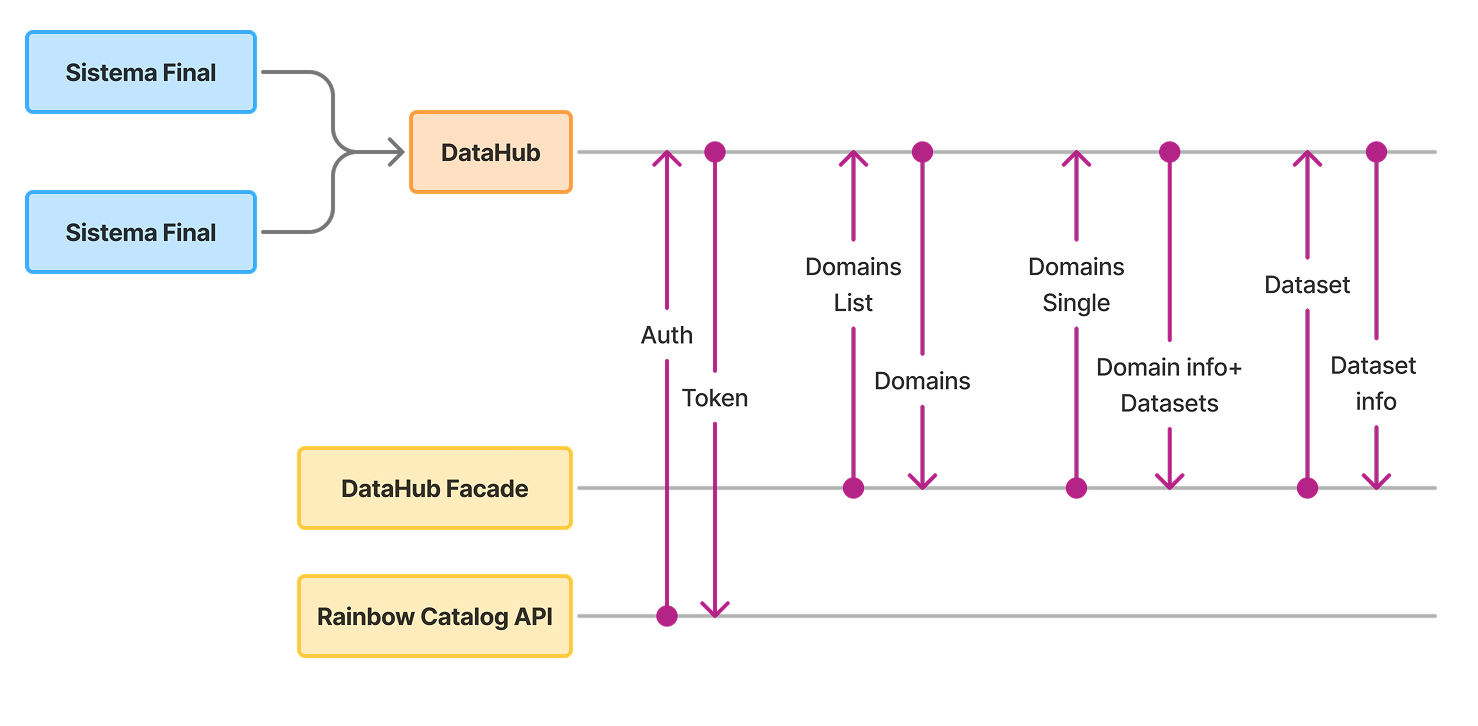}
}
\caption{Flujo de consulta de metadatos.}
\label{fig:protocols}
\end{figure*}

\subsection{Flujo de exploración y selección de \textit{target}}
Este flujo arranca en Rainbow mediante una fachada que actúa como pasarela contra DataHub, como se aprecia en la Figura 4. Primero, la fachada se autentica y obtiene un token. Con ese contexto, emite consultas al servidor GraphQL habilitado en DataHub para listar \textit{domains} que se proyectan como catálogos a nivel de DCAT y, después, recuperar el detalle de un \textit{domain} junto a los \textit{datasets} que contiene.

Para cada \textit{dataset} seleccionado se solicita su detalle operativo, extrayendo la identidad única (el URN canónico de DataHub) y la información de distribución necesaria para los flujos del \textit{Dataspace Protocol}: tipo de distribución, puntos de acceso (\textit{endpoints}) y posibles configuraciones del servicio de acceso/consumo (p. ej., autenticación), así como más metadatos definidos. El URN canónico de DataHub se emplea como atributo \textit{target} de la política ODRL. 

Con esta información, Rainbow mapea el modelo interno de DataHub a DCAT y publica una vista DCAT neutra (\textit{endpoint} del \textit{Dataspace Provider}), de modo que los participantes del espacio de datos pueden consultar catálogos y \textit{datasets} sin conocer el modelo de entidades de DataHub.

\begin{figure*}[htbp] 
\centerline{
\includegraphics[width=0.76\textwidth]{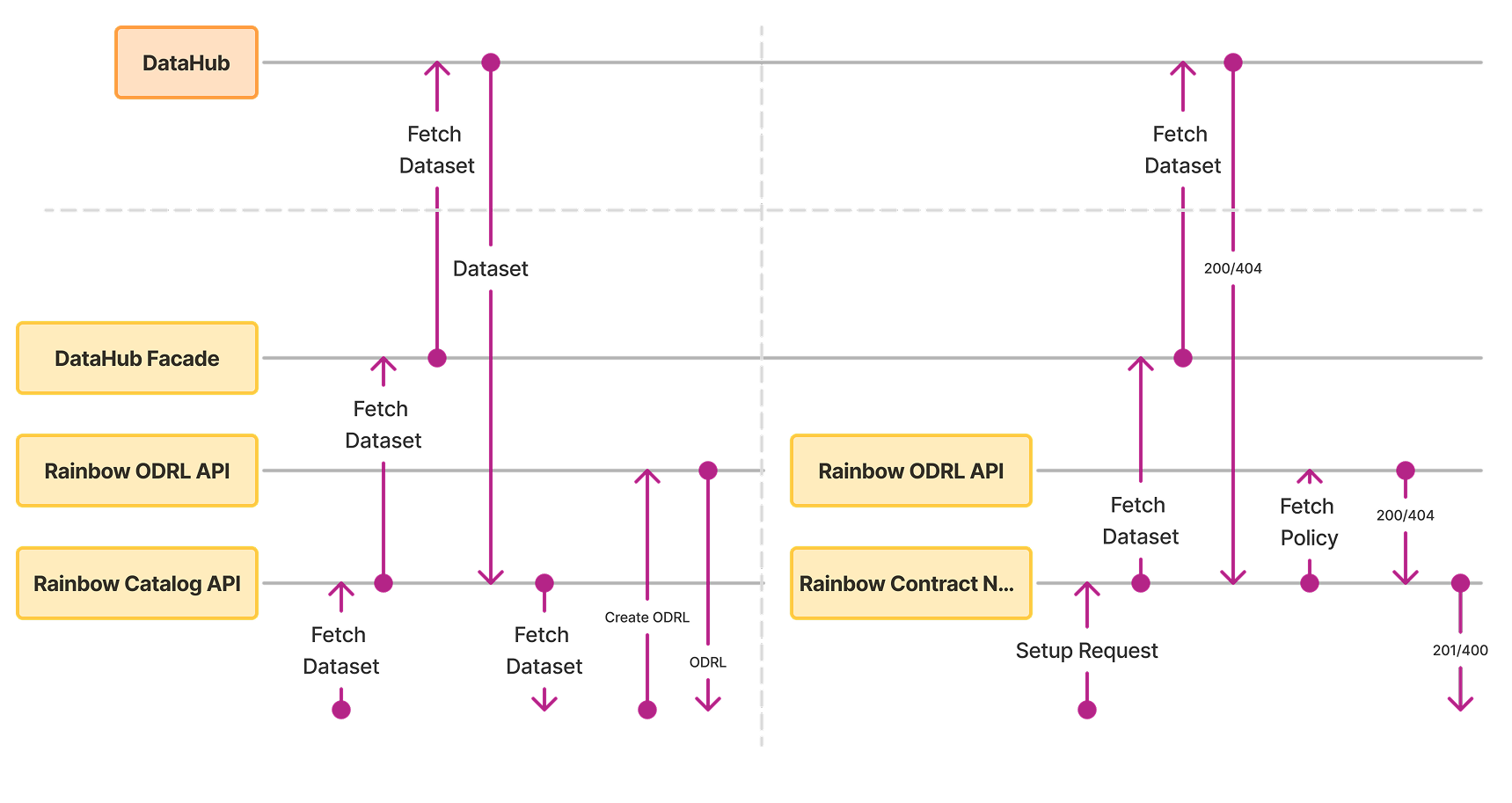}
}
\caption{Flujo de creación y aplicación de políticas}
\label{fig:protocols}
\end{figure*}

\subsection{Flujo de creación y aplicación de políticas}
Con el \textit{dataset} identificado en el flujo anterior, Rainbow Catalog gestiona políticas ODRL de forma nativa como se ilustra en la Figura 5. El ciclo es:
\begin{enumerate}
    \item Consulta/validación del \textit{dataset}: antes de asociar una política, la Rainbow ODRL API pide a la fachada \textit{dataset} para verificar que el \textit{dataset} existe y que los metadatos operativos (distribución, \textit{endpoint}) son consistentes.
    \item Creación y persistencia de la política: se conforma la política ODRL del \textit{dataset} en DataHub y se persiste en Rainbow con su propio identificador.
\item Negociación de contrato (CN): cuando el \textit{Dataspace Consumer} inicia la negociación, Rainbow realiza una doble comprobación: vuelve a consultar el \textit{dataset} vía fachada y recupera la política. De ser aceptada por ambas partes se genera un \textit{Agreement} con el contenido ODRL de dicha política.
\item Negociación de transferencia (TP): usando el \textit{target} de la política almacenado en el \textit{Agreement}, Rainbow resuelve de nuevo, a través de la fachada, el tipo de distribución, el punto de acceso y las configuraciones pertinentes del sistema final. Con ello autoriza y entrega al consumidor el acceso conforme a las condiciones pactadas.

\end{enumerate}

Con este esquema en dos fases, DataHub aporta gobernanza, linaje, ingesta y federación a gran escala, mientras que Rainbow aporta exposición DCAT interoperable y gestión de políticas ODRL estrechamente acoplada a los flujos de negociación de contrato y negociación de transferencia del \textit{Dataspace Protocol}. El resultado es una integración coherente: descubrimos y resolvemos metadatos operativos (fase 1) y negociamos/transferimos con control de uso (fase 2), siempre anclados al URN del \textit{dataset} federado.

\subsection{Aplicación al caso de uso}

Sobre el caso de uso, se ha aplicado todo el sistema de ETL desde los catálogos de datos para extraer los metadatos con Apache NiFi, la orquestación de con Apache Airflow de manera exitosa. Se puede, por tanto, generar una instancia federadora y dos instancias federadas con DataHub. El repositorio se puede encontrar en este enlace: \url{https://github.com/pvinualese/datahub\_dockerized}.

Por otro lado, se lanzó una instancia de Rainbow, donde se generó un módulo ad-hoc para operar como \textit{proxy} contra DataHub. Dicho flujo se encuentra en el notebook bajo este enlace: \url{https://github.com/ging/ds-deployment/blob/main/src/pull\_flow\_rpc\_datahub.ipynb}. Como se puede observar en el flujo, se usa el \textit{proxy} para acceder a los datos de los \textit{datasets} de DataHub, y posteriormente se pueden persistir y usar políticas ODRL asociadas al \textit{dataset} y persistirlas en Rainbow. 

\subsection{Desafíos y trabajos a futuro}
Durante la implementación, han aparecido retos técnicos que se tratará de abordar en futuros trabajos, tales como: 
\begin{itemize}
    \item Coherencia y sincronización de metadatos: En caso de haber cambios en DataHub como modificaciones o bajas en los mismos, las políticas quedarían invalidadas al no poder resolver \textit{targets}. Una opción posible sería valorar sistemas de suscripción a cambios y definir una interfaz para ello. 
    \item Rendimiento en consultas federadas: En caso de haber un número alto de \textit{datasets} puede que el sistema de consultas basado en GraphQL actúe como cuello de botella. Una posible solución sea la de consultas estrechas a DataHub (sólo con datos necesarios), paginaciones o preindexaciones y sistemas de caché en la fachada de Rainbow. 
    \item Mapeo semántico de DataHub a DCAT. DataHub tiene un sistema de entidades propio y el mapeo a DCAT no es trivial. Tanto como para la gestión de campos complejos como ODRL, como para campos relacionales como sistemas de versiones, ect.
    \item Gestión de herencia y ámbito de políticas. Actualmente las políticas se asocian con \textit{datasets}, pero se debería investigar en asociar las políticas a nivel de Catálogo, u otras entidades definidas en DCAT. Para ello haría falta una redefinición del uso de políticas a nivel de \textit{Dataspace Protocol} que pretendemos abordar en trabajos futuros. 
\end{itemize}

\section{\uppercase{Conclusiones}}
Este documento se ha centrado en hacer una propuesta de implementación sobre una cuestión crucial dentro de los ecosistemas de espacios de datos y su posible adopción masiva por parte de actores proveedores de servicios de datos que es la integración de catálogos federados que puedan funcionar a gran escala y que estén basados en soluciones existentes y probadas en la industria, y al mismo tiempo que puedan abordar los requisitos de interoperabilidad inherentes a los espacios de datos.

En dicho documento se ha hecho una revisión del papel que juega el catálogo centrándonos en los flujos de negociación y transferencia. Por otro lado se ha hecho una comparativa sobre una implementación existente y desarrollada por el grupo de investigación que propone este artículo, que parte de los principios de interoperabilidad y uso de estándares, muy cercana al caso de uso relacionado con los flujos de un espacio de datos; y por otro lado de una solución como DataHub con gran adopción a nivel global desde el punto de vista de gestión y gobernanza del dato. 

En la investigación hemos detectado la necesidad de hacer una implementación híbrida, definiendo las capas de ingesta, gestión y persistencia de metadatos bajo DataHub, y definiendo las capas de creación, adjudicación, gestión y persistencia de las políticas ODRL en Rainbow. Entendemos que es una solución que desde el punto de vista de arquitectura es coherente, y además en la implementación práctica está resultando exitosa.

Esta integración es total y abre la puerta al desarrollo de una interfaz de usuario, desde la cual se podrán hacer ofertas para acceder a un recurso, negociar las condiciones de acceso. De esta manera, consumidores y proveedores podrán llegar a acuerdos de cara permitir la transferencia de los datos. Al mismo tiempo se recalca que implementaciones de catálogos federados permiten escalar el impacto de los espacios de datos a casos de uso masivos en gestión de metadatos y de gestión de la gobernanza desde un punto de vista integral.

\section*{\uppercase{Agradecimientos}}

Los autores agradecen el apoyo del proyecto EUNOMIA, "Soluciones para la soberanía, confianza y Seguridad en los espacios de datos" (ETD202300196), financiado por el INCIBE (Instituto de Ciberseguridad).\url{https://eunomia.dit.upm.es}, y de la Agencia Estatal de Investigación (AEI), 10.13039/501100011033., a través del proyecto FuN4Date con la subvención PID2022-136684OB-C22.

\bibliographystyle{IEEEtran}

\IEEEpeerreviewmaketitle

\end{document}